\input epsfig.sty
\documentstyle[12pt]{article}
\begin{document}

\def\PRL#1{{\it Phys.~Rev.~Lett.~}{\bf #1}}
\def\PRD#1{{\it Phys.~Rev.~}{\bf D#1}}
\def\PR#1{{\it Phys.~Rev.~}{\bf #1}}
\def\NPB#1{{\it Nucl.~Phys.~}{\bf B#1}}
\def\NP#1{{\it Nucl.~Phys.~}{\bf #1}}
\def\PLB#1{{\it Phys.~Lett.~}{\bf B#1}}
\def\ARNPS#1{{\it Ann.~Rev.~Nucl.~Part.~Sci.~}{\bf #1}}
\def\ZPhysC#1{{\it Z.~Phys.~}{\bf C#1}}
\def\PRepC#1{{\it Phys.~Rep.~}{\bf C#1}}
\def\ProgTP#1{{\it Prog.~Th.~Phys.~}{\bf #1}}
\def\ModPL#1{{\it Mod.~Phys.~Lett.~}{\bf A#1}}

\def\beq{\begin{equation}}
\def\eeq{\end{equation}}
\def\beqa{\begin{eqnarray}}
\def\eeqa{\end{eqnarray}}
\def\sm{Standard Model~}
\def\ps{$SU(4)\times SU(2)_L\times SU(2)_R$~}
\def\cf{{\it cf.} }
\def\ie{{\it i.e.} }
\def\Slash{\hskip -.6em/}
\def\ul{\underline}

\def\la{~\mbox{\raisebox{-.6ex}{$\stackrel{<}{\sim}$}}~}
\def\ga{~\mbox{\raisebox{-.6ex}{$\stackrel{>}{\sim}$}}~}

\def\Slash#1{%
   \setbox0=\hbox{$#1$}\dimen0=\wd0\setbox1=\hbox{/}\dimen1=\wd1%
   \ifdim\dimen0>\dimen1\rlap{\hbox to \dimen0{\hfil/\hfil}}#1%
   \else\rlap{\hbox to \dimen1{\hfil$#1$\hfil}}/\fi}%

\rightline{EFI 95-43}
\bigskip
\rightline{August 1995}
\bigskip
\bigskip
\renewcommand{\thefootnote}{\fnsymbol{footnote}}

{\centerline{\bf COSMOLOGICAL BARYON NUMBER AND KAON CP VIOLATION}}
{\centerline {\bf FROM A COMMON SOURCE.}}

\bigskip
\centerline{\it Mihir P. Worah\footnote{
Address after October 1: Stanford Linear Accelerator Center, Stanford
University, Stanford, CA 94309}}
\smallskip
\centerline{Enrico Fermi Institute and Department of Physics}
\centerline{University of Chicago, Chicago, IL 60637}
\bigskip
\centerline{\bf Abstract}

We extend an earlier model for radiatively generated fermion masses
based on the Pati-Salam group to include CP violation. Spontaneous CP
violation in the early universe gives rise to a complex mass matrix
for heavy sterile neutrinos. The out-of-equilibrium decay of these
neutrinos generates a $B-L$ asymmetry. The sterile neutrinos also act as
a mass seed in generating one-loop (complex) mass matrices for the
quarks. Thus, the two low energy manifestations of CP violation -- the
CKM phase and the baryon number asymmetry -- can both be traced in a
calculable way to a common source.

\newpage

\section{Introduction.}
\renewcommand{\thefootnote}{\arabic{footnote}}
\setcounter{footnote}{0}
\medskip

All of the information we have about CP violation can be summarized by
two seemingly unrelated numbers: $\epsilon = 2.26\pm 0.02 \times
10^{-3}$ and
$\eta_B\equiv n_B/n_{\gamma} = (2.8-4.0)\times 10^{-10}$ \cite{pdg}.
These two numbers are a
measure of two seemingly unrelated phenomena: the strange decays of the
$K$ mesons and the baryon asymmetry of the universe.
It is generally agreed upon that the best candidate for the source of
$\epsilon$ is an unremovable phase in the quark mixing (CKM)
matrix \cite{ckm}; however, there is no such agreement on a single most
likely source of $\eta_B$.

Although there are a number of viable models of baryogenesis
most of them do not say anything about the CKM phase, a notable
exception to this being some of the papers that propose that the
baryon asymmetry is generated
at the electroweak phase transition with the only source of CP
violation being in the quark mass matrix \cite{ewb}. Although
electroweak baryogenesis remains an exciting possibility \cite{ckn}
the general consensus is
that additional sources of CP violation are needed to make it work
\cite{noewb}.

An interesting mechanism of generating the baryon asymmetry is to
first generate a lepton asymmetry by the out-of-equilibrium decay
of a heavy sterile neutrino and then use the $B+L$ anomaly in the
\sm to cycle this into a baryon asymmetry \cite{fy1,fy2}.
This scenario fits naturally into a model for radiatively generated
fermion masses based on the Pati-Salam
gauge group \ps \cite{ps} that we have presented
\cite{mw}. The motivation for this model was firstly to treat the
quarks and leptons on a symmetric footing, and then to show that one can
generate realistic masses and mixings for the \sm fermions without a
large hierarchy of Yukawa couplings.

In this model the \sm fermions (including right-handed partners for
the neutrinos) are
supplemented by three generations of heavy sterile neutrinos.
As a result of a particularly simple choice of scalar representations
it is impossible for the \sm fermions to get masses at tree-level,
and the sterile neutrinos act as mass seeds in generating finite
one-loop masses for them. If we allow the sterile neutrinos to have a
complex mass matrix (as the result of a spontaneous violation of
CP in the early universe), then not only does their out-of-equilibrium
decay generate a
baryon number asymmetry, but their exchange in the one-loop diagrams
also generates a complex mass matrix for the quarks. Thus one is able
to derive both $\epsilon$ and $\eta_B$ from the same source: a complex
phase in the sterile neutrino mass matrix that gets communicated to
the quark sector at the scale where lepton number breaks down. It is the
purpose of this paper to explore this possibility.

In a sense this model is complementary to the models of electroweak
baryogenesis where one starts with the CKM phase
and tries to use it to generate a fermion number
asymmetry. Here we start with a
fermion number violating operator and try
to use it to generate a phase in the CKM matrix.

In Sec.~2 we briefly review the model, its representation content and
pattern of symmetry breakdown. In Sec.~3
we calculate the baryon asymmetry generated in this model in two
alternative scenarios: one in which there was a period of inflation in
the early universe, and one in which there wasn't. In Sec.~4 we
discuss the fermion masses and CKM matrix, and illustrate with a
numerical example. We conclude in Sec.~5.
Some details of the scalar potential and explicit formulas for the
fermion masses are presented in Appendices A and B.

\section{The Model}

\medskip

The gauge group is \ps with gauge couplings $g_S,~g_L$ and $g_R$.
The \sm fermions transform in the usual
representations:
\beq
\Psi_L^i\sim   ({\ul 4},{\ul 2},{\ul 1})^i\equiv
             \left(\begin{array}{cccc} u_1& u_2& u_3& \nu \\
                           d_1& d_2& d_3& e^- \end{array}\right)_L^i
\eeq
\beq
\Psi_R^i\sim   ({\ul 4},{\ul 1},{\ul 2})^i\equiv
             \left(\begin{array}{cccc} u_1& u_2& u_3& N \\
                           d_1& d_2& d_3& e^- \end{array}\right)_R^i
\eeq
where $i=1,2,3$ is a generation index, and we have included a right
handed neutrino $N$. We add to this three
generations of (right-handed) sterile neutrinos
\beq
s^i \sim  ({\ul 1},{\ul 1},{\ul 1})^i.
\eeq
We choose to make the matter spectrum supersymmetric,
{\footnote{This differs from the scalar spectrum in Ref.~\cite{mw}
where we had only two generations of $L$ and $R$, and no $\sigma$. For
other related models see Refs.~\cite{rabi1,volkas}.}}
so the scalars in the model transform as
\beq
L^i\sim   ({\ul 4},{\ul 2},{\ul 1})^i \equiv
          \left(\begin{array}{cccc} L_{u1}& L_{u2}& L_{u3}& L_{\nu} \\
                           L_{d1}& L_{d2}& L_{d3}& L_{e} \end{array}\right)^i,
\eeq
\beq
R^i\sim   ({\ul 4},{\ul 1},{\ul 2})^i \equiv
          \left(\begin{array}{cccc} R_{u1}& R_{u2}& R_{u3}& R_{N} \\
                           R_{d1}& R_{d2}& R_{d3}& R_{e} \end{array}\right)^i,
\eeq
and
\beq
\sigma^i \sim ({\ul 1},{\ul 1},{\ul 1})^i.
\eeq
We will impose a discrete $Z_3$ symmetry on the gauge singlets
(broken by the interactions of the \sm particles) under
which $s^j \rightarrow e^{-i(j\pi)/3} s^j$ and $\sigma^j \rightarrow
e^{i(2j\pi)/3}\sigma^j$. This permits us to make the Lagrangian CP
invariant, with the vacuum expectation values of the $\sigma^j$
breaking CP spontaneously as in Ref.~\cite{branco}.

The Yukawa interactions will then be
\beq
{\cal L_{Y}} = -y_i(\bar s^c)^i s^i \sigma^i
               -(\kappa_L^a)_{ij}\bar \Psi_L^i s^j L^a
               -(\kappa_R^a)^T_{ij}\bar \Psi_R^i (s^c)^j R^a + {\rm h.c.},
\label{gaugeyuk}
\eeq
with all of the coupling constants real.
We discuss the details of the scalar potential in Appendix A, noting
only that we can choose parameters such that it
is minimized when
\beq
\langle\sigma\rangle_j = \frac{v_0}{\sqrt 2}e^{i\alpha_j};~~~
\langle R_N\rangle_j = \frac{v_R}{\sqrt 2}\delta_{1j};~~~
\langle L_{\nu}\rangle_j = \frac{v_L}{\sqrt 2}\delta_{1j}
\label{vevs}
\eeq
with $|v_0| > |v_R| \gg |v_L|$.
None of the \sm fermions get masses at tree level. Their masses are
generated at one-loop by diagrams involving the sterile neutrinos
$(s_i)$ on internal lines. The pattern of
symmetry breaking induced by the scalar vacuum expectation values is
\beqa
SU(4)_C\times SU(2)_L\times SU(2)_R\times CP
 &\buildrel v_0\over\longrightarrow & SU(4)_C\times SU(2)_L\times
SU(2)_R \nonumber \\
 &\buildrel v_R\over\longrightarrow & SU(3)_C\times SU(2)_L\times
U(1)_Y \nonumber \\
 &\buildrel v_L\over\longrightarrow & SU(3)_C\times U(1)_Q.
\label{symbrk}
\eeqa
If we use as inputs at the electroweak scale $\alpha^{-1}=128.5$,
$s_W^2=0.23$, and $\alpha_S^{-1}=8.33$, and the boundary condition
$g_L(v_R) = g_R(v_R)$, then using the one-loop
$\beta$ functions with only gauge boson and fermion contributions,
gives us $v_R = 10^{14}$ GeV. The masses of the sterile neutrinos are
required to be close to $v_R$ in order to optimize the radiative mass
generation. Thus we choose $v_0 = 10^{15}$ GeV.
All of the scalars have a mass $\sim v_R$ except for $L_{\nu}^1$
and $L_e^1$ which remain light and form the components of the
\sm Higgs boson.

\section{Baryogenesis}

In this section we work in the symmetric phase of the Standard Model,
{\em i.e.}, $v_L=0$. We shall denote the light scalar doublet with
components $(L_{\nu}^1,L_e^1)$ as $\Phi^*$, the (complex conjugated)
\sm Higgs boson. The out-of-equilibrium decay that we are then
concerned with is that of the
lightest (right-handed) $SU(2)\times U(1)$ singlet neutrino $n_1$
into a left-handed lepton,
$l_L^i$, and the Higgs boson, $\Phi$ \cite{fy1}.
In order to calculate this, we first need to re-express the Yukawa
interactions of Fig.~1(a) involving the singlet neutrinos
$s_i$ $(i=1,2,3)$ in terms of the physical (Majorana)
neutrino states $n_j$
($j=1,...,6$) obtained when one diagonalizes the neutrino mass matrix
[Fig.~1(b)].

\epsfig{file=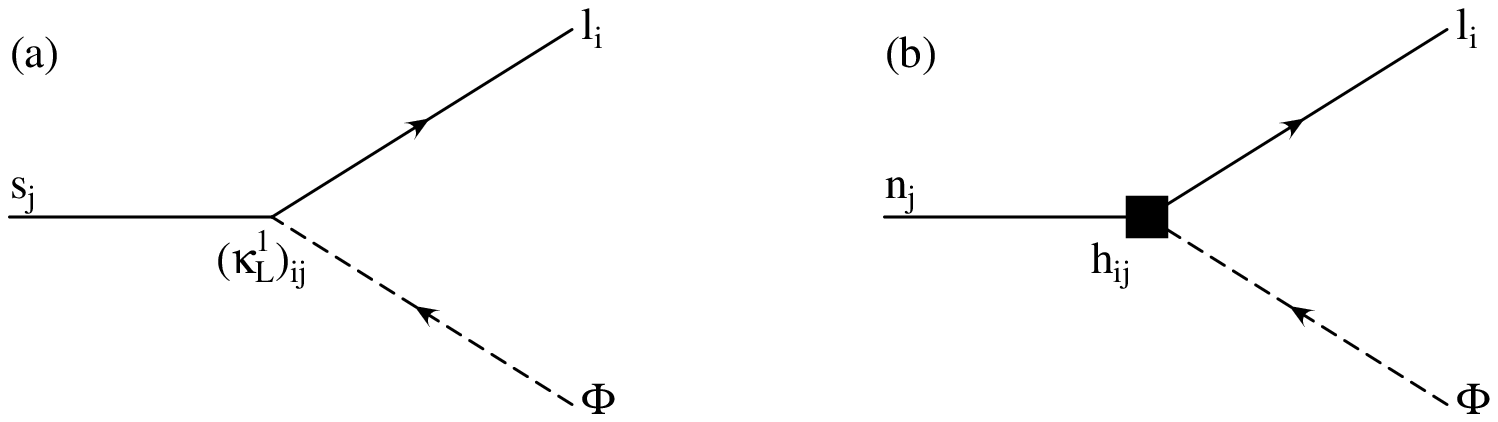,width=6in}
{\small {\bf Fig. 1} Yukawa couplings of the heavy singlet
neutrinos to a light left-handed lepton doublet and the \sm Higgs. The
arrows indicate flow of left-handed isospin; $(\kappa_L^1)_{ij}$ and
$h_{ij}$ are coupling constants.
(a) Coupling of the $s_j$. (b) Coupling of the physical states $n_j$.}

\vspace{0.5cm}

The vacuum expectation values of Eq.~(\ref{vevs})
along with the Yukawa couplings of Eq.~(\ref{gaugeyuk})
lead to the following $6\times 6$ mass matrix for the right-handed
neutrinos in the basis $(N_i,s_i)$ with $i=1,2,3$:
\beq
M =
\left(\begin{tabular}{c|c}
$0$ & ${\displaystyle{\frac{(\kappa_R^1) v_R}{2\sqrt 2}}}$ \\ \hline
${\displaystyle{\frac{(\kappa_R^1)^T v_R}{2\sqrt 2}}}$ &
$\begin{array}{ccc}
m_{01} e^{i\alpha_1} & 0 & 0 \\
0 & m_{02} e^{i\alpha_2} & 0 \\
0 & 0 & m_{03} e^{i\alpha_3}\end{array}$
\end{tabular}\right)
\eeq
where $m_{0i}=y_i v_0/\sqrt 2$. The matrix $M$ will be diagonalized by a
$6\times 6$ unitary matrix $U$ with $M = U M_D U^T$ where $M_D$ is
diagonal, real and positive.

We can now obtain the Yukawa couplings of
the physical heavy neutrinos, $n^j$, with the light left-handed
leptons, $l^i$, and the Higgs boson $\Phi$:
\beq
{\cal L}_Y' = -h_{ij}\bar l_L^i n_R^j \Phi + {\rm h.c.}
\label{effyuk}
\eeq
where
\beq
h_{ij} = \sum_{k=1}^3 (\kappa_L^1)_{ik}U_{k+3,j}^*
\label{yukis}
\eeq
is now a complex $3\times 6$ matrix, and $n_j$ are the 6 physical
right-handed neutrinos arranged so that $m_1 < m_2 < ... < m_6$.

{}From here the calculation proceeds just as the cases presented in the
literature (except for the fact that we have to account for 6 heavy
neutrinos rather than 3), depending on whether we assume a period of
inflation in the early universe \cite{olive} or not \cite{fy1,luty},
and we now illustrate with a numerical example that the model can
naturally generate a baryon asymmetry of the correct size.

We choose
as inputs the following matrices of Yukawa couplings
(we will motivate this choice in the next section):
\beq
 \begin{array}{cc}
\kappa_L^1=\left(\begin{array}{ccc}
                 0.04& 0.03& 0.06\\
                 0.06& 0.42& 0.24\\
		 0.06& 0.08& 3.5\end{array}\right);
&\kappa_R^1=\left(\begin{array}{ccc}
                 -0.06& 0.03& -0.04\\
                 0.06& 0.20& 0.16\\
		 0.06& 0.08& 3.5\end{array}\right)
\end{array}
\label{yukin1}
\eeq
\beq
y_1 = \sqrt 2\times 0.025;~~~y_2 = \sqrt 2\times 0.05;~~~
y_3=\sqrt 2\times 0.35.
\label{yukin2}
\eeq
We make a simple choice for the phases $\alpha_i$ of
Eq.~(\ref{vevs}):
\beq
-{\pi} < \alpha_1 < {\pi};~~~\alpha_2=\alpha_3=0.
\label{phases}
\eeq

Since $n_1$ is a Majorana
fermion it can decay via the Yukawa interaction of Eq.~(\ref{effyuk})
into both leptons and anti-leptons, {\em i.e.},
\beqa
n_1 & \rightarrow & l_L^i + \Phi \nonumber \\
    & \rightarrow & (l_L^c)^i + \Phi^*,
\eeqa
thus violating lepton number in its decays. The tree level width of
$n_1$ is then
\beq
\Gamma_t = \frac{(h^{\dagger}h)_{11} m_1}{8\pi}
\label{tree}
\eeq
where $h$ is defined in Eq.~(\ref{yukis}) and
\beq
(h^{\dagger}h)_{11}=3\times 10^{-5},
\label{hdhnum}
\eeq
and
\beq
m_1 = 4 \times 10^{11}~GeV
\label{m1}
\eeq
for the inputs  we have chosen.

Further, if CP is
violated there will be an asymmetry in the decay rates into leptons
and anti-leptons. Let us define an asymmetry
\beq
\delta = \frac{\Gamma - \Gamma^{CP}}{\Gamma + \Gamma^{CP}}
\label{defdelta}
\eeq
where $\Gamma$ is the decay rate into leptons, and $\Gamma^{CP}$ into
anti-leptons. Interference between the diagrams of Figs.~2(a) and 2(b)
give rise to $\delta$ which is calculated to be~\cite{fy1,luty}
\beq
\delta = \frac{1}{2\pi(h^{\dagger}h)_{11}}
         \sum_{j=1}^6 {\rm Im}[(h^{\dagger}h)_{1j}]^2
                                      f(m_j^2/m_1^2),
\label{delta}
\eeq
where
\beq
f(x) = \sqrt{x} \left[1-(1+x) {\rm ln}\left(\frac{1+x}{x}\right)\right]
\eeq
is proportional to the imaginary part of the loop amplitude of
Fig.~2(b), and arises when the $l_k$ and $\Phi$ in the loop go on-shell.

\epsfig{file=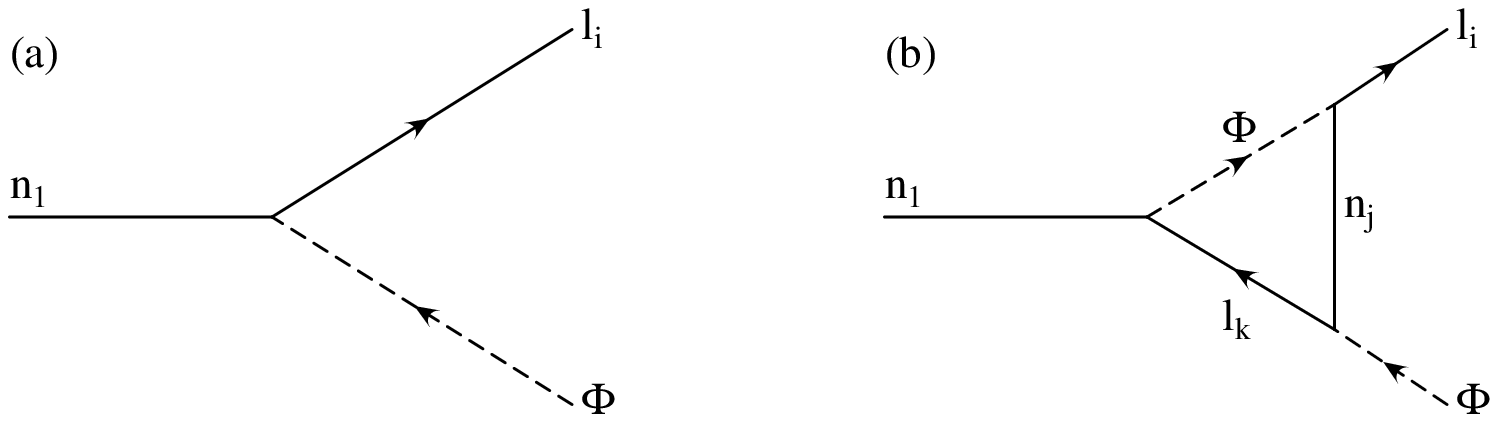,width=6in}
{\small {\bf Fig. 2}. Decay of the lightest physical singlet
neutrino $n_1$ into a left-handed lepton and the \sm Higgs.
(a) Tree level. (b) One-loop, with on-shell intermediate states.}

\vspace{0.5cm}

The calculation now splits up depending on whether we want to
incorporate the effects of inflation or not.
We will first estimate the baryon asymmetry
in the inflationary case (although this calculation is
more speculative than the non-inflationary one).
We will then adapt the result to the case of no inflation.
{\footnote{For a text book discussion of some of the estimates we make
in this section and their accuracy, see Ref.~\cite{kolbturner}.}}

Quantum fluctuations of the scalar field $\eta$ (the inflaton),
that drives inflation, give rise to density perturbations in the early
universe which in turn lead to
anisotropies in the cosmic microwave background radiation.
Thus one can use the COBE measurements of the quadrupole
moment of the microwave background \cite{cobe} to constrain the
parameters of inflation. For a generic inflationary potential as
defined in Ref.~\cite{olive}, one finds
\beq
\frac{\mu}{M_P}\sim 10^{-4}
\label{infscale}
\eeq
where $\mu$ is the inflation scale, and $M_P=10^{19}$ GeV is the
Planck mass. Thus inflation takes place at a scale $\mu\sim 10^{15}$
GeV.

The fact that inflation occurs at or below $v_0=10^{15}$ GeV, the scale at
which CP is spontaneously broken, solves the problem of
domain walls in this model. A domain
wall arises whenever a discrete symmetry is spontaneously broken, and
contains an unacceptably large energy density \cite{okun}. If one
existed in our Hubble volume, it would have over-closed the universe
many times over \cite{langacker}.
Inflation at a scale below that of the discrete symmetry breakdown has
the effect of diluting the density of domain walls to the level where
it is extremely unlikely to find one in our observable universe.

The mass of the inflaton is
\beq
m_{\eta} \simeq \frac{\mu^2}{M_P}
          \simeq 10^{11}{\rm GeV}
\label{minfl}
\eeq
and the reheat temperature is
\beq
T_{RH}\simeq (\Gamma_{\eta} M_P)^{1/2}
       \simeq \frac{\mu^3}{M_P^2}
        \simeq 10^7-10^8 ~{\rm GeV}
\label{trh}
\eeq
where $\Gamma_{\eta}\simeq m_{\eta}^3/M_P^2$ is the inflaton
decay rate, and $T_{RH}$ is defined as the equilibrium temperature of
the relativistic decay products of the inflaton.

Thus $m_{\eta} \simeq 10^{11}$ GeV $ \ga m_1$, and
the inflaton is heavy enough to decay into the right-handed neutrino
$n_1$ whose subsequent decay generates a lepton asymmetry.
A simple way to estimate the baryon asymmetry in this case is \cite{olive}
\beqa
\frac{n_L}{s} &\simeq &\frac{n_{\eta}}{s}\delta
              \simeq \frac{\rho_{\eta}}{m_{\eta} s}\delta \nonumber \\
              &\simeq &\frac{T_{RH}}{m_{\eta}}\delta
                \sim (10^{-3}-10^{-4})\delta.
\eeqa
Here we have used the approximation that all of inflaton's energy
density, $\rho_{\eta}$, is instantaneously converted
into the energy density of relativistic particles: $\rho_{\eta}\simeq
g^{*} T_{RH}^4$. The subsequent entropy density is $s\simeq g^{*}
T_{RH}^3$ ($g^*$ is the number of relativistic degrees of
freedom), and  $m_{\eta}$ and $T_{RH}$ are obtained from Eq.~(\ref{minfl})
and Eq.~(\ref{trh}) above.
The baryon asymmetry is then
\beq
\eta_B \equiv \frac{n_B}{n_{\gamma}} = -\frac{28}{79}\times
              7\frac{n_L}{s} \sim -(10^{-3}-10^{-4})\delta
\eeq
where we have used
\beq
\frac{n_B}{s}=\frac{28}{79}\frac{n_{(B-L)}}{s}
\eeq
as a result of the electroweak $B+L$ violation \cite{ht},
and the fact that $s=7n_{\gamma}$ today.
In Fig.~3 we plot the CP asymmetry $\delta$ as a function of the phase
$\alpha_1$ for the choice of inputs of Eqs.~(\ref{yukin1}, \ref{yukin2}).
We see that the choice $\alpha_1 = -\pi/2$ gives us
$\delta=-3 \times 10^{-7}$
and subsequently $\eta_B \simeq 10^{-10}$ in agreement with the
observed value. We pick the
value with $|\alpha_1|=\pi/2$ as it might be consistent with
the notion of a ``maximal'' CP violation in the early universe.

\begin{center}
\epsfig{file=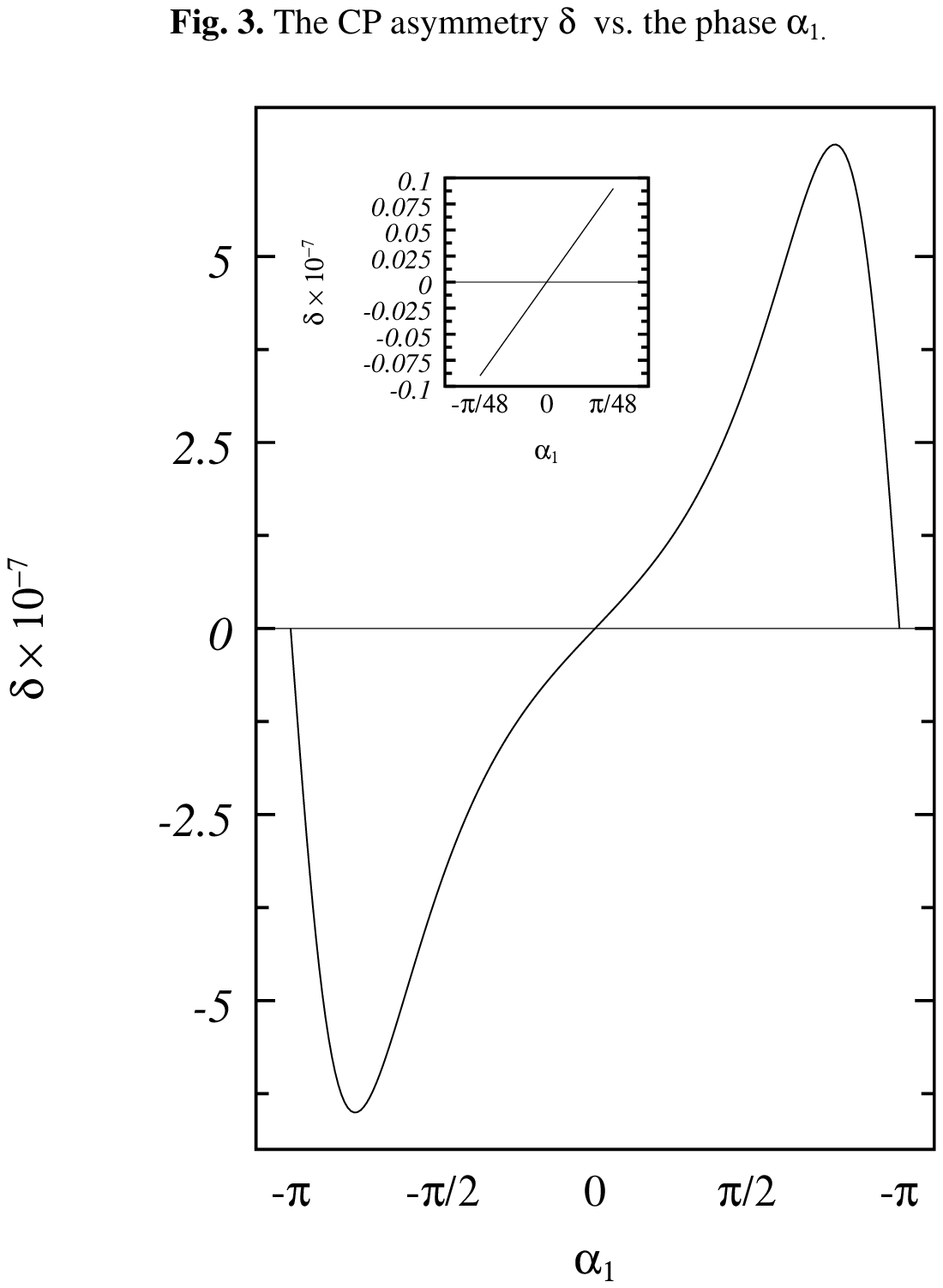, width=6in}
\end{center}

The last thing we need to ensure is that the lepton
asymmetry is not washed out by the lepton number violating process
$l_L l_L\rightarrow \Phi \Phi$ of Fig.~4.

\begin{center}
\epsfig{file=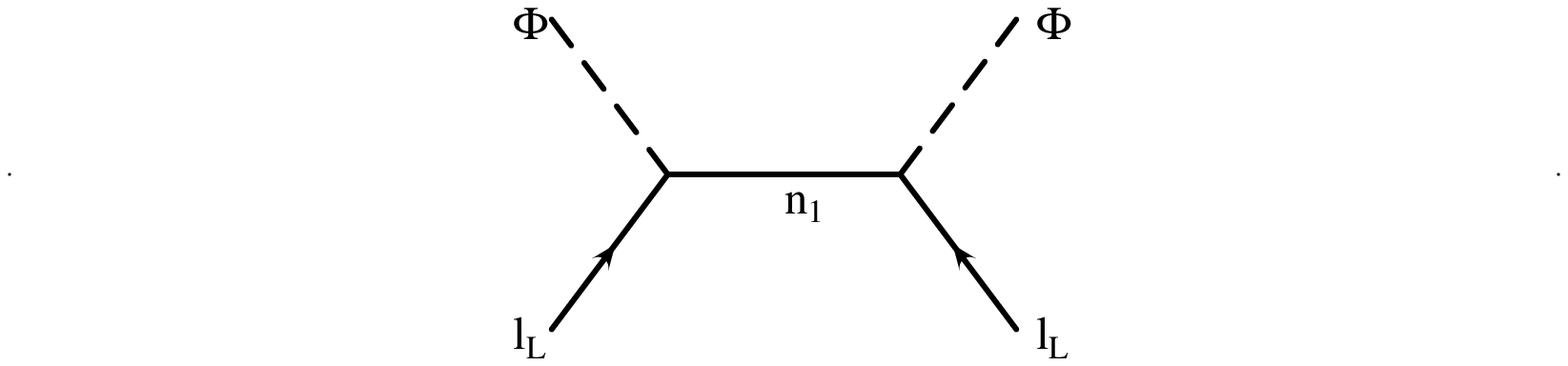,width=6in}
{\small {\bf Fig. 4.} Induced $\Delta L=2$ interaction. The arrows
indicate flow of lepton number.}
\end{center}

\vspace{0.5cm}
This requires that the process be out of equilibrium when the
electroweak $B+L$ violation is in equilibrium \cite{fy2,olive,ht}, which
happens at a temperature $\tilde{T}_R \sim \alpha^2 T_{RH} \sim 10^5$ GeV,
where $\alpha$ is a typical gauge coupling \cite{blah}.
The bound one needs to satisfy is
\beq
\frac{m_1}{(h^{\dagger}h)_{11}} \ge 1.4 \times 10^{-2}~\sqrt{\tilde{T}_R
M_P}
\label{washout}
\eeq
which is easily satisfied for the parameters we use.

Although we prefer the inflationary scenario because it cures the
domain wall problem as mentioned above, and because of its other
attractive features, we now show how to adapt the
calculation to the case of no inflation.

In the non-inflationary case, we first need to check if $n_1$ is out
of equilibrium when it decays {\em i.e.}
\beq
\Gamma_t  \le  H(T=m_1)
\eeq
where $H$ is the Hubble constant. This translates to
\beq
\frac{(h^{\dagger}h)_{11}m_1}{8\pi} \la \frac{20 m_1^2}{M_P}
\eeq
which is satisfied for our choice of inputs.
The lepton asymmetry is then given by
\beqa
\frac{n_L}{s} & \simeq & \frac{n_{\gamma}}{s}\delta
		\simeq \frac{n_{\gamma}}{g_*n_{\gamma}} \nonumber \\
              & \simeq & 10^{-2}\delta
\eeqa
where we have used $g^*\simeq 100$ between the scales $v_L$ and $v_R$.
Thus we see that in this case a sufficient baryon asymmetry can be
achieved with a smaller CP violation than in the inflationary
case. This is borne out by Fig. 3 where we see that smaller values of
$\alpha_1$ indeed lead to smaller values of $\delta$. If we choose
$\alpha_1 = -\pi/96$, we get $\delta = -3 \times 10^{-9}$ and
$\eta_B\simeq 10^{-10}$.

Once again we need to ensure against a wash-out of this symmetry. In
this case, the $B+L$ violation is in equilibrium up to a temperature of
about $10^{12}$ GeV, and hence in the RHS of Eq.~({\ref{washout}) we
use the temperature at which the lepton number violation occurs:
$T_L=m_1=10^{11}$ GeV. The inequality
\beq
\frac{m_1}{(h^{\dagger}h)_{11}} \ge 1.4 \times 10^{-2}~\sqrt{T_L
M_P}
\eeq
is indeed satisfied, and once again we find that our model can
generate a sufficient baryon asymmetry.

\section{Fermion Masses and the CKM matrix}

The calculation for the fermion masses and mixings exactly follows
that in \cite{mw}, except that there are now 3 generations of scalars
rather than 2, and that we now allow for CP violation, hence the
orthogonal matrix $O$ in the mass formulas
is replaced by the unitary matrix $U$, or $U^*$ as appropriate.
We relegate the details to Appendix B,
restricting the discussion in this section to qualitative
features, and demonstrating with a numerical example.

As mentioned earlier, none of the \sm fermions get masses at tree
level, even after the standard electroweak group $SU(2)_L\times
U(1)_Y$ is broken. However, one-loop masses are generated by the
diagrams of Figs.~(5). $X$ in Fig.~5(a) is an $SU(4)$ lepto-quark
gauge boson, and $Z'$ of Figs.~5(e),~5(f) is the neutral gauge boson
that couples to the linear combination of the broken diagonal
generators $U(1)_{B-L}$ and
$U(1)_R$ of $SU(4)$ and $SU(2)_R$ respectively, that is orthogonal to
the hypercharge $U(1)_Y$. $Z$ of Fig.~5(f) is
just the \sm $Z$ boson.

One can estimate the magnitude of the mass terms from Figs.~5(a) and 5(e)
involving gauge boson exchange as
\beq
m_{G1} \simeq \frac{g_1g_2}{(4\pi)^2} v_L  v_R
           \frac{m_0}{m_0^2}
    \simeq \frac{g_1 g_2}{(4\pi)^2} v_L
\label{mgauge}
\eeq
where $g_1$ and $g_2$ are the gauge couplings at the two vertices. The
factor of $(4\pi)^2$ in the denominator is from the loop integral,
and we have replaced the sterile neutrino mass $m_0$ by $v_R$.

\newpage

\epsfig{file=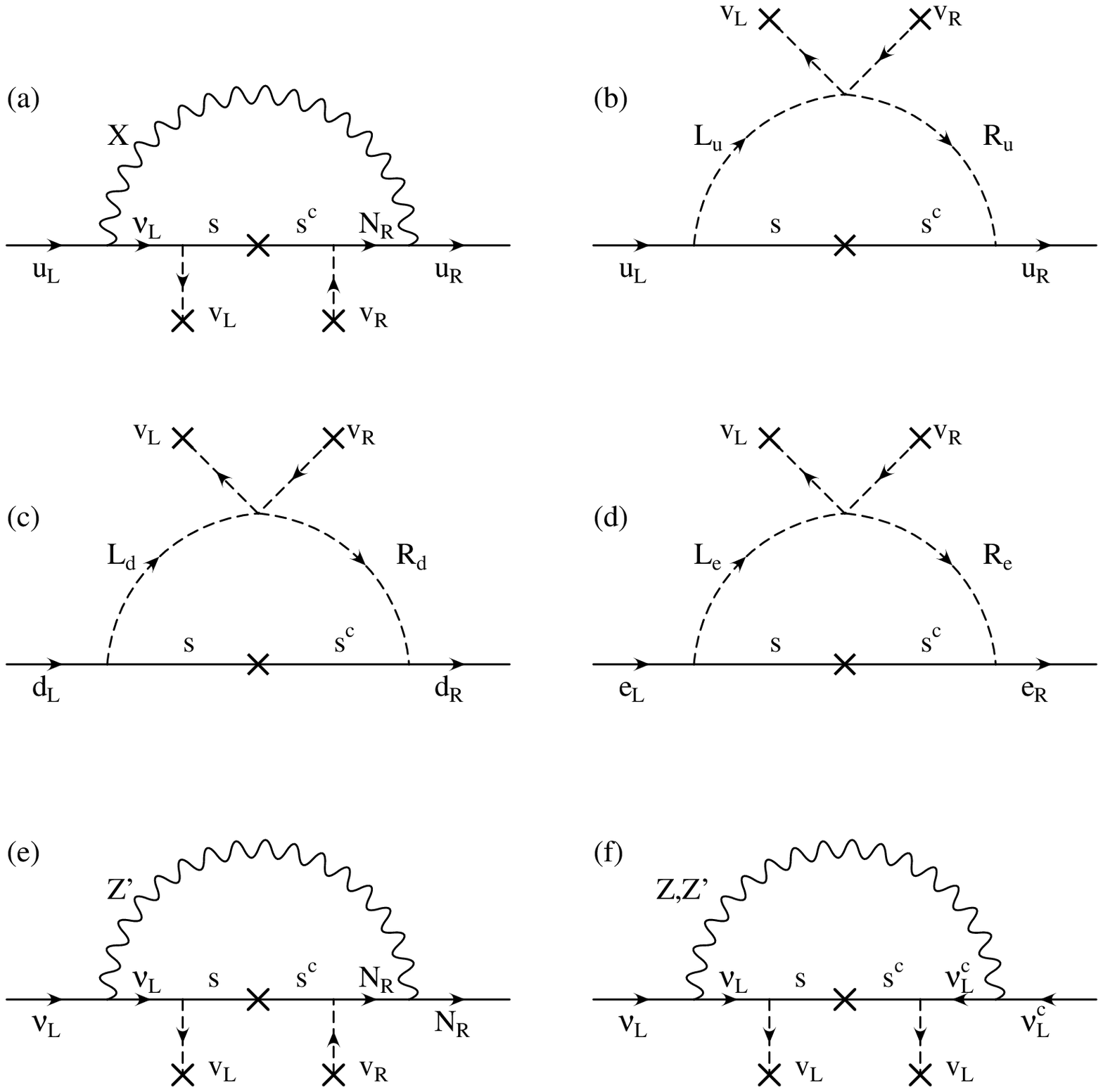,width=6in}
\vspace{0.2cm}
{\small {\bf Fig. 5.} One-loop processes that generate the fermion masses.
The arrows here indicate the flow of left or right-handed isospin.
(a) Up-type quarks with gauge bosons. (b) Up-type quarks with
scalars.
(c) Down-type quarks. (d) Charged leptons.
(e) Dirac mass for neutrinos. (f). Majorana mass for neutrinos.}

\newpage

Notice that if we replace $v_L$ on the external leg of Fig.~5(e) by
the unshifted Higgs field, $\Phi$, we get an effective interaction
similar to that of Fig.~1(a), but with $N_j$ on the external leg. One
might then worry about whether it was correct to ignore this
contribution to the baryogenesis calculation of the previous
section. To estimate the magnitude of the effective Yukawa couplings
of the $N_j$ to the $l_i$ and $\Phi$,
one can simply divide the result of Eq.~(\ref{mgauge}) by $v_L$. If we
use $g_1g_2/(4\pi) \simeq 1/40$, we get effective Yukawa couplings
$y \simeq 0.002$ which are smaller than the
couplings $\kappa_L^1$ of Eq.~(\ref{yukin1}) by an order of
magnitude. Thus their
contribution to the total width of $n_1$ can be be ignored.
The fact that this process makes
no contribution to the CP asymmetry can be understood by
replacing all the particles in Fig.~5(e) by physical particles. Then,
with an on-shell $n_1$ on the external leg, and $n_i$ and $Z'$ in the
loop, the diagram can never have an imaginary part since $n_1$ is the
lightest physical neutrino, and the particles in the loop cannot be
on-shell. These estimates were confirmed by a direct calculation of
the appropriate three-point form factors, and their contributions
to the baryogenesis calculation.

Using arguments similar to the ones for Figs. 5(a) and 5(e), one can
estimate Fig.~5(f) to give
\beq
m_{G2} \simeq\frac{g_1 g_2}{(4\pi)^2}\frac{v_L}{v_R} v_L,
\label{mmaj}
\eeq
and the scalar exchange contribution of Figs.~5(b), 5(c) and 5(d)
to the fermion masses are approximately given by
\beq
m_S\simeq \sum_{a=1}^3\frac{\kappa_L^a\kappa_R^a}{(4\pi)^2} v_L
\label{mscal}
\eeq

One immediately notices a fundamental
difference between fermion masses in this model and those in the
Standard Model. Here all the fermion masses are proportional to
the squares of
coupling constants as opposed to depending linearly on them.{\footnote
{A more careful estimate would show that the gauge
contribution of Eq.~(\ref{mgauge}) is numerically similar to the
scalar contribution with $\kappa_L^1$ and $\kappa_R^1$ of
Eq.~(\ref{mscal}). This must be since some of the scalars $L^1$ and
$R^1$ form the longitudinal components of the massive gauge bosons,
{\cf} Eqs.~(\ref{vevs},\ref{symbrk}).
The choice of ``gauge'' or
``scalar'' contribution then depends on the gauge we calculate in,
with the final answer being gauge independent.}}
This has the effect of reducing the hierarchy in coupling constants
needed to reproduce the large hierarchy in the observed masses.

If this model were supersymmetric then the estimates for the fermion
masses of the previous equations would be scaled by a factor of
$M_{SUSY}/v_R$ for $M_{SUSY} < v_R$,
where $M_{SUSY}$ is the supersymmetry breaking scale.
Thus the discovery of TeV scale supersymmetry
would rule out this model for generating the fermion masses.

Although all the quarks and leptons of one generation have identical
Yukawa couplings in this model, the scalars to which they couple have
different masses and mixings determined by the parameters of the
scalar potential (the explicit formulas for the masses are listed in
Appendix B). Using this fact, and some of the features of
Figs.~5, one can qualitatively understand how
this model gives rise to the rather complicated observed spectrum of
fermion masses.

The up-type quarks get masses from both gauge boson and scalar
exchange [Figs.~5(a),~5(b)].
This larger number of diagrams, and the possibility of
constructive or destructive interference between them allows us to
generate the large hierarchy in their masses. In particular all the
diagrams that contribute to the top quark mass interfere
constructively to generate the large observed value, whereas
destructive interference between various diagrams is responsible for
making the up quark lighter than the down quark.

The down-type quarks and charged leptons both get masses only from
scalar exchange, explaining the similarity in their spectra
[Figs.~5(c),~5(d)].

The neutrinos get masses from gauge boson exchange [Figs.~5(e),~5(f)].
However, the large
mass of the sterile neutrinos results in see-saw suppression of the
physical left-handed neutrino mass to a level compatible with the MSW
\cite{msw} solution to the solar neutrino problem. The neutrinos could
in principle also get masses from scalar exchange, but this
contribution vanishes for a natural choice of parameters in the scalar
potential.

The complex phase in the CKM matrix arises as a result of the single phase
$\alpha_1$ in the sterile neutrino mass matrix.

Let us illustrate these features with a numerical example.
We choose the following Yukawa couplings in addition to
$\kappa_L^1,~\kappa_R^1,~y_i$ of
Eqs.~(\ref{yukin1}, \ref{yukin2}),
\beq
\begin{array}{cc}
\kappa_L^2=\left(\begin{array}{ccc}
                 0.04& 0.03& 0.06\\
                 0.03& 0.4& 0.2\\
		 0.1&  3.5& 3.5\end{array}\right);
&\kappa_R^2=\left(\begin{array}{ccc}
                 0.04& 0.03& 0.06\\
                 -0.03& 0.4& -0.2\\
		 0.1&  3.5& 3.5\end{array}\right)\\
\kappa_L^3=\left(\begin{array}{ccc}
                 -0.2& 0.02& -0.2\\
                 -0.2& 0.4&  -0.4\\
		 3.5& 3.5& 3.5\end{array}\right);
&\kappa_R^3=\left(\begin{array}{ccc}
                 0.2& -0.02& 0.2\\
                 -0.4& -2.0& -1.0\\
		 3.5& 3.5& 3.5\end{array}\right)
\end{array}
\label{yukin3}
\eeq

The procedure we used to pick the Yukawa couplings was the
following. The diagonal entries of $\kappa_{L,R}^{(1,2)}$ were
arbitrarily chosen to
reflect the hierarchy $1:\lambda:\lambda^2$ with $\lambda = 0.1$,
going from the third generation to the first,
in accordance with our policy of not allowing too large a
hierarchy in the couplings. The scale was set by the $(3,3)$ entry
which was picked to be $3.5$, the largest value with which we felt
comfortable doing perturbation theory. The first row and column were
chosen to reflect the magnitude of the (1,1) element, and the (2,3)
and (3,2) elements were arbitrarily chosen to reflect either the (2,2)
or the (3,3) elements. Some of the entries were randomly
assigned minus signs (except the entries in the third row, which
contribute to the top quark mass, and were all required to be
positive). For the $\kappa_{L,R}^3$, all the elements of the third row
were set to 3.5 in order to maximize the top quark mass, and the
elements of the first and second rows were varied in order to
fit the data.

Given the inputs above, for a choice of parameters of the scalar
potential similar to those of \cite{mw} one obtains the following
fermion masses and mixings (at the electroweak scale $\sim 100$ GeV).
\beq
m_u=7.0~MeV;~~~m_c=1.2~GeV;~~~m_t=160~GeV.
\label{upqEW}
\eeq
\beq
m_d=8.0~MeV;~~~m_s=140~MeV;~~~m_b=3.9~GeV.
\label{dnqEW}
\eeq
The absolute values of the CKM matrix elements are
\beq
|V|=\left(\begin{array}{ccc}
                 0.98& 0.21& 0.004\\
                 0.21& 0.98& 0.06\\
		 0.01& 0.06& 1.0\end{array}\right)
\label{Vkm}
\eeq

As a measure of the CP violation in the CKM matrix we first calculate
the parameter $J_{CP}$, where
\beq
J_{CP} = {\rm Im}[V_{cb}V_{us}V_{ub}^*V_{cs}^*]
\label{jcp}
\eeq
is twice the area of the unitarity triangle of CKM
matrix elements, and is a rephasing invariant measure of the CP
violation in the CKM matrix with 3 generations of quarks.
We plot $J_{CP}$ as a function of the phase $\alpha_1$ in Fig.~6.
In the Wolfenstein parametrization of the CKM matrix \cite{wolfi}, we
have $J_{CP}=A^2\lambda^6\eta\simeq 4\times 10^{-5}$ where we have
used $A=0.85,~\lambda=0.22$, and $\eta\simeq 0.5$.
For our
choice of inputs, and with $\alpha_1=-\pi/2$ we see from Fig.~6 that
\beq
J_{CP}=3 \times 10^{-5},
\label{jcp2}
\eeq
which has the correct magnitude and sign.

Notice that if we attach a photon to the charged particles on the
internal lines of the diagrams in Fig.~5, we could induce an electric
dipole moment for the quarks (and charged leptons)
at one-loop order. This is to be
contrasted with the Standard Model, where the fermion electric dipole
moments arise only at three-loop order \cite{donoghue,shab1}.
An estimate of the dipole moment is then
\beq
d_{\psi} \sim \frac{m_{\psi}}{v_R^2}~{\rm e \cdot cm}
\label{edm}
\eeq
where $m_{\psi}$ is the mass of the fermion in question. This is well
below the estimated \sm value for the quarks \cite{shab2}, and hence
the low energy hadronic CP violation in this model is effectively
of the CKM type.

\begin{center}
\epsfig{file=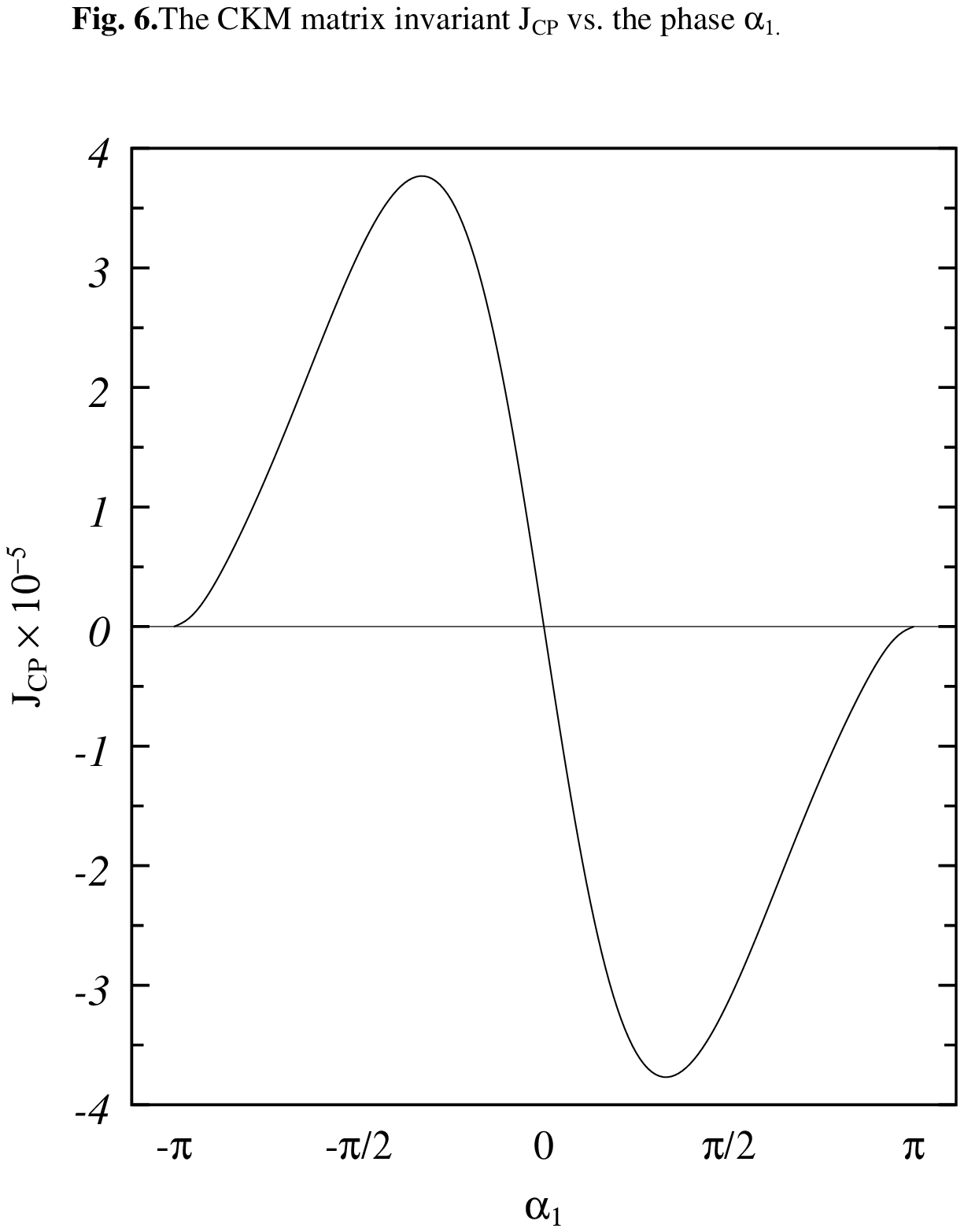, width=6in}
\end{center}

In order to estimate the kaon CP violating parameter $\epsilon$, we
assume that the top quark contribution dominates the box diagram that
leads to $K^0-\bar K^0$ mixing. In this case one has \cite{rosner,burhas}
\beqa
|\epsilon|& \simeq C_{\epsilon}B_KJ_{CP}|V_{ts}|^2 \nonumber \\
          &\simeq 3 \times 10^{-3}
\eeqa
where we have used
\beq
C_{\epsilon}=\frac{G_F^2F_K^2m_KM_W^2}{6\sqrt 2\pi^2\Delta M}
            = 3.8 \times 10^4,
\eeq
and $B_K=0.65$ ($B_K$ measures the accuracy of the vacuum saturation
approximation).

In the lepton sector we obtain
\beq
m_e=0.7~MeV;~~~m_{\mu}=90~MeV;~~~m_{\tau}=1.3~GeV.
\label{elm}
\eeq
\beq
m_{\nu_e}=7\times 10^{-5}~eV;~~~m_{\nu_{\mu}}=4\times 10^{-3}~eV;~~~
m_{\nu_{\tau}}=3\times 10^{-2}~eV.
\label{nem}
\eeq
The absolute values of the lepton mixing matrix are
\beq
|V_{\nu}|=\left(\begin{array}{ccc}
                 0.92& 0.39& 0.03\\
                 0.39& 0.91& 0.12\\
		 0.03& 0.12& 0.99\end{array}\right)
\label{Vlep}
\eeq
We should point out that since the neutrino masses depend sensitively
on the scale $v_R$ via the see-saw mechanism, the above values should
be taken as representative of the range the actual masses should
lie in. This prediction of the scale of the neutrino masses is then
fairly robust since they are generated only by gauge boson exchange,
and hence independent of the Yukawa matrices $\kappa_{L,R}^{(2,3)}$
and details of the scalar potential
(they do depend on $\kappa_{L,R}^1$ which, however, are
constrained by the baryogenesis calculation of Sec.~3).
It is interesting to note, then, that the $\nu_e$ and
$\nu_{\mu}$ masses lie in the correct range to explain the solar
neutrino deficit as a result of the MSW \cite{msw} effect, with the
mixing angle being compatible with the large angle solution \cite{sankar}.
We should point out, however, that the prediction for this
mixing angle is not as robust as that of the masses since
it is dominated by the contribution of the unitary matrix that
diagonalizes the charged lepton mass matrix, and hence dependent on
details of the scalar sector.
Although the $\nu_{\mu}-\nu_{\tau}$ mass difference
lies in the correct range for the proposed solution to the atmospheric
neutrino problem, the required mixing angle is generically much
smaller than that required by the theoretical fits to the data
\cite{sankar}.

This model also predicts CP violation in the lepton sector,
due to complex phases in the lepton mixing matrix.
However, these effects are suppressed by the smallness of the
neutrino masses, and hence, unobservable. If one applies the estimate
of Eq.~(\ref{edm}) for the electric dipole momments of the charged
leptons, one gets a value larger than what one would calculate just
using the phases in the mixing matrix \cite{donoghue}, however it is
still much to small to be observed.

Thus we are able to generate realistic fermion masses and
mixings within our stated objective of working in a quark-lepton
symmetric model without a large hierarchy
of coupling constants. Not only that, but starting with the same phase
$\alpha_1 = -\pi/2$ in the sterile neutrino mass matrix that generates
a lepton (and hence baryon) asymmetry, we are also able to generate
the correct amount of CP violation in the CKM matrix Eq.~(\ref{jcp2}).

All of the inputs to the model were defined at the high energy scale
$v_R=10^{14}$ GeV. We would like to remark on the procedure we have
followed in obtaining the outputs at $100$ GeV. The effective theory
below the scale $v_R$ is just the \sm and hence we run the masses
using \sm $\beta$ functions. Effectively this results in a scaling
of the light quark
masses by a factor of 2.5 to account for the QCD
effects, whereas the top quark mass scales by a factor of 1.7, on account
of its slower running due to its large Yukawa coupling to the \sm
Higgs.
None of the mixing angles, nor the lepton masses run to the degree of
accuracy we're concerned with.

As we have mentioned, the discovery of low energy supersymmetry would
rule out this mechanism of mass generation. However the scenario
whereby a CP violating mass matrix for sterile neutrinos is the source
of both the baryon asymmetry and the CKM phase may still remain
viable. We are currently studying the possibility of building this
scenario into a supersymmetric model. Another possibility we are
studying is the extension of this model to include family symmetries
for the \sm fermions in order to reduce the arbitrariness of the
Yukawa couplings. The fact that the Yukawa couplings in this model
only range from ${\cal O}(1)$ to ${\cal O}(\lambda^2)$ with $\lambda =
0.1$, may allow a simple assignment of family charge to the \sm
fermions, and the possible use of non-abelian family
symmetries \cite{seibergramond}.

\section{Conclusions}

We have presented a simple extension of the \sm based on the
Pati-Salam gauge group $SU(4)\times SU(2)_L\times SU(2)_R$.
CP is spontaneously broken in the early universe, and the CP
violation manifests itself as a single phase of $-\pi/2$ in the mass
matrix of the sterile neutrinos that are present in our model. With
this phase as the source, we can generate both the baryon asymmetry of
the universe, $\eta_B\simeq 10^{-10}$, and $\epsilon \simeq 10^{-3}$,
the CP violating asymmetry in kaon decays.

Masses for the fermions are generated
radiatively at one-loop order, and we can obtain realistic masses and
mixings without the large hierarchy of coupling constants required in
the Standard Model.
The neutrino masses lie in the correct range for the MSW solution to
the solar neutrino problem. There will be CKM-type CP violation in the
lepton sector, however the smallness of the neutrino masses makes any
effects unobservable.

\section{Acknowledgements}

The author would like to acknowledge useful discussions with Aaron
Grant, Jerry Jungman, Alex Kagan and Jon Rosner. He would especially
like to thank Aaron for sharing his numerical routine to calculate
3 point functions, and Jon for valuable comments on the manuscript.
This work was supported in part by the United States
Deparment of Energy under Grant No. DE-FG02-90ER40560.

\bigskip
\bigskip

{\Large{\bf Appendix A: The Scalar Potential}}

\medskip
\medskip

\noindent Given the hierarchy $v_0>v_R\gg v_L$, the simplest way
to study the scalar potential is to analyze the three
sectors comprising $\sigma^i$, $R^i$, and $L^i$ independently. While
this suffers from a lack of generality, it makes the analysis
extremely transparent without affecting our results.

For the $\sigma^i$ we work with a simplified version of the $Z_3$
invariant scalar potential of Ref.~\cite{branco}.
\beqa
V(\sigma) &= -2m_1^2(\sigma_1^{\dagger}\sigma_1)
              -2m_2^2(\sigma_2^{\dagger}\sigma_2)
              -2m_3^2(\sigma_3^{\dagger}\sigma_3) \nonumber \\
           &  +a_1(\sigma_1^{\dagger}\sigma_1)^2
              +a_2(\sigma_2^{\dagger}\sigma_2)^2
              +a_3(\sigma_3^{\dagger}\sigma_3)^2 \nonumber \\
  & +[C_1(\sigma_1^{\dagger}\sigma_2)(\sigma_1^{\dagger}\sigma_3)
    +C_2(\sigma_2^{\dagger}\sigma_1)(\sigma_2^{\dagger}\sigma_3)
    +C_3(\sigma_3^{\dagger}\sigma_1)(\sigma_3^{\dagger}\sigma_2) +
{\rm h.c.}]
\eeqa
We then assume the vacuum expectation value $\langle \sigma\rangle_j =
v_je^{i\alpha_j}/\sqrt 2$, and study the equations to minimize the
potential. In particular, for the choice $C_2=C_3=0,~m_i\simeq
10^{15}$ GeV, we recover
\beq
v_1^2=v_2^2=v_3^2=v_0^2=\frac{m_1^2}{a_1-C_1};~~
\alpha_1=-\frac{\pi}{2},~\alpha_2=\alpha_3=0
\label{vev0}
\eeq
which are the values used in Eqs.~(\ref{vevs}, \ref{phases}).

For the $R^i$ and $L^i$, we specialize to the
one-family case, since $L^1\equiv L$ and $R^1\equiv R$ are the only
scalars that get vacuum expectation values (Eq.~(\ref{vevs})).
\beq
V(R)=-2\mu_R^2 R_{i\alpha}R^{i\alpha} +
\lambda_{R1}R_{i\alpha}R^{i\alpha} R_{j\beta}R^{j\beta} +
\lambda_{R2}R_{i\alpha}R^{j\alpha} R^{i\beta} R_{j\beta} +
\lambda_{R3}R_{i\alpha}R^{j\alpha} R^{i}_{\beta}R_{j}^{\beta}
\eeq
\beq
V(L)=-2\mu_L^2 L_{i\alpha}L^{i\alpha} +
\lambda_{L1}L_{i\alpha}L^{i\alpha} L_{j\beta}L^{j\beta} +
\lambda_{L2}L_{i\alpha}L^{j\alpha} L^{i\beta} L_{j\beta} +
\lambda_{L3}L_{i\alpha}L^{j\alpha} L^{i}_{\beta}L_{j}^{\beta}.
\eeq
Here $i=1,2$ is the $SU(2)_L$ or $SU(2)_R$ index, and $\alpha=1,2,3,4$
is the $SU(4)$ index. $L^{i\alpha}=(L_{i\alpha})^*$ and
$L^{i}_{\alpha}=\epsilon^{ij}L_{j\alpha} $, and similarly for
$L\rightarrow R$. We can then generalize the arguments of
Ref.~\cite{li} to show that if the following conditions are satisfied:
\beqa
\lambda_{R2} &< 0;~~\lambda_{R1}+\lambda_{R2} >
0;~~\lambda_{R3}>\lambda_{R2}~~ or~~
|\lambda_{R3}|>2\lambda_{R1}+\lambda_{R2} \nonumber \\
\lambda_{L2} &< 0;~~\lambda_{L1}+\lambda_{L2} >
0;~~\lambda_{L3}>\lambda_{L2}~~or~~|\lambda_{L3}|>2\lambda_{L1}+\lambda_{L2}
\eeqa
the absolute minimum of the potential is at
\beqa
\langle R\rangle &=\left(\begin{array}{cccc} 0& 0& 0& v_R/\sqrt 2 \\
                           0& 0& 0& 0 \end{array}\right) \nonumber \\
\langle L\rangle &=\left(\begin{array}{cccc} 0& 0& 0& v_L/\sqrt 2 \\
                           0& 0& 0& 0 \end{array}\right)
\label{vevlr}
\eeqa
with
\beqa
(\lambda_{R1}+\lambda_{R2})v_R^2&=&2\mu_R^2 \nonumber \\
(\lambda_{L1}+\lambda_{L2})v_L^2&=&2\mu_L^2
\label{minlr}.
\eeqa
For $\mu_R\simeq 10^{14}$ GeV and $\mu_L\simeq 100$ GeV, we then
obtain $v_R\simeq 10^{14}$ GeV and $v_L\simeq 100$ GeV.
There will be other constraints on the parameters of the scalar
potential to ensure $\langle L\rangle^j=\langle R\rangle^j=0$ for
$j\ne 1$, which are easily satisfied for natural choices of the
parameters.
The only light scalars are then
$L_{\nu}^1$ and $L_e^1$, with the rest having masses $\sim v_R$.

We would like to point out that in
order for scalar exchange to generate fermion masses one actually
needs $L^i-R^i$ mixing to interchange left and right weak
isospin as one can see from Figs.~5(b),~5(c),~5(d).
What we do in the actual calculation is to allow individual
couplings to be non-zero, but require that the combination of
constants that couple $v_L$ to $v_R$ in the equations that determine
the minimum to add up to zero, thus leaving Eq.~(\ref{minlr})
unchanged even in the more general case (see Ref.~\cite{mw} for
details).

Another point we would like to make is that one could in principle
start with a scalar potential that was $L-R$ symmetric, and use
the couplings of the $L^i,~R^i$ to the $\sigma^i$ to break this
symmetry as in Ref.~\cite{rabi}.

\bigskip
\bigskip
\newpage

{\Large{\bf Appendix B: Details on the Fermion Masses}}

\medskip
\medskip

\noindent
In this appendix we would like to fill in some of the details,
and list the explicit formulas used to calculate the fermion masses.

All the calculations were done at zero external momentum, and
with physical particles propagating in
the loop diagrams of Fig.~(5). Thus, before we get to the formulas for
calculating the \sm fermion masses, we need to obtain the masses and
mixing angles for the singlet neutrinos, gauge bosons, and scalars
that propagate in the the diagrams of Fig.~(5).

The vacuum expectation values for the scalars, $\sigma^i$, $L^1$, and
$R^1$ [Eqs.~(\ref{vevs}, \ref{vev0}, \ref{vevlr})], along with the
Yukawa couplings of Eq.~(\ref{gaugeyuk}) generate the following $9
\times 9$ mass matrix for the neutrinos in the basis
$(\nu_i^c,N_i,s_i)$ for $i=1,2,3$,
\beq
M =
\left(\begin{tabular}{c|c|c}
$0$ & $0$ &
${\displaystyle{\frac{(\kappa_L^1) v_L}{2\sqrt 2}}}$ \\ \hline
$0$ & $0$ &
${\displaystyle{\frac{(\kappa_R^1) v_R}{2\sqrt 2}}}$ \\ \hline
${\displaystyle{\frac{(\kappa_L^1)^T v_L}{2\sqrt 2}}}$ &
${\displaystyle{\frac{(\kappa_R^1)^T v_R}{2\sqrt 2}}}$ &
$\begin{array}{ccc}
m_{01} e^{i\alpha_1} & 0 & 0 \\
0 & m_{02} e^{i\alpha_2} & 0 \\
0 & 0 & m_{03} e^{i\alpha_3}\end{array}$
\end{tabular}\right)
\label{n9mass}
\eeq
where $m_{0i}=y_i v_0/\sqrt 2$. The matrix $M$ will be diagonalized by a
$9\times 9$ unitary matrix $U$ with $M = U M_D U^T$ where $M_D$ is
diagonal, real and positive. This matrix has eigenvalues
$m_1 = m_2 = m_3 = 0$ corresponding to the physical states
$n_1,~n_2,~n_3$ that are mostly admixtures of the $\nu_i$, and
$m_4,...,m_9 \simeq v_R$ for the physical states $n_4,...,n_9$
that are mostly admixtures of the $N_i$ and $s_i$. The mixing between
the $(\nu_i)$ and the $(N_i,~s_i)$ is of order $v_L/v_R$.

The gauge bosons for the gauge group $SU(4)
\times SU(2)_L \times SU(2)_R$ are:
{\small
\beq
\hat G_{\mu}  = \frac{1}{2}
\left(\begin{array}{cccc} G_{3\mu}+
{\displaystyle{\frac{G_{8\mu}}{\sqrt 3}+\frac{B_{\mu}}{\sqrt 6}}}
& \sqrt 2 G_{12\mu}^+& \sqrt 2 G_{13\mu}^+& \sqrt 2 X_{1\mu}^+\\
\sqrt 2 G_{12\mu}^-&
-G_{3\mu}+
{\displaystyle{\frac{G_{8\mu}}{\sqrt 3}+\frac{B_{\mu}}{\sqrt 6}}}
& \sqrt 2 G_{23\mu}^+& \sqrt 2 X_{2\mu}^+\\
\sqrt 2 G_{13\mu}^-& \sqrt 2 G_{23\mu}^-&
{\displaystyle{-\frac{2G_{8\mu}}{\sqrt 3}+\frac{B_{\mu}}{\sqrt 6}}}
& \sqrt 2 X_{3\mu}^+\\
\sqrt 2 X_{1\mu}^-& \sqrt 2 X_{2\mu}^-& \sqrt 2 X_{3\mu}^-&
{\displaystyle{-\frac{3 B_{\mu}}{\sqrt 6}}} \end{array}\right),
\eeq
\beq
\hat W_{L\mu}  = \frac{1}{2}
                \left(\begin{array}{cc} W_{L\mu}^0& \sqrt 2 W_{L\mu}^+ \\
                \sqrt 2 W_{L\mu}^-& -W_{L\mu}^0 \end{array}\right),
\eeq
}
and
{\small
\beq
\hat W_{R\mu} = \frac{1}{2}
                \left(\begin{array}{cc} W_{R\mu}^0& \sqrt 2 W_{R\mu}^+ \\
                \sqrt 2 W_{R\mu}^-& -W_{R\mu}^0 \end{array}\right),
\eeq
}
where the $G_{\mu}$ are the gluons, $B_{\mu}$ is the diagonal
gauge boson that couples to
$B-L$, and the $X_{\mu}$ are the lepto-quarks.

As a result of the spontaneous symmetry breaking outlined in
Eq.~(\ref{symbrk}) the charged gauge bosons get the following masses:
\beq
M_X^2
=\frac{g_S^2}{4}[v_R^2+v_L^2];~~M_{W_L}^2=\frac{g_L^2}
{4} v_L^2;~~M_{W_R}^2=\frac{g_R^2}{4} v_R^2.~~~~~~~~~~~~
\label{cbmass}
\eeq
The neutral gauge bosons have the mass squared matrix
\beq
M_{0} = \frac{1}{8}\left(\begin{array}{ccc}
                   g_L^2v_L^2& 0& -(3/\sqrt 6)g_Lg_Sv_L^2 \\
                    0& g_R^2v_R^2& -(3/\sqrt 6)g_Rg_Sv_R^2 \\
               -(3/\sqrt 6)g_Lg_Sv_L^2& -(3/\sqrt 6)g_Rg_Sv_R^2&
                (3/2)g_S^2(v_R^2+v_L^2)
                            \end{array}\right)
\eeq
in the basis ($W_L^0,W_R^0,B^0$), with eigenvalues
\beq
M_{\gamma}^2=0;~M_Z^2=\frac{v_L^2}{4}[g_L^2+\frac{3g_R^2g_S^2}
{2g_R^2+3g_S^2}+{\cal O}(v_L^2/v_R^2)];
{}~M_{Z'}^2=\frac{v_R^2}{8}[2g_R^2+3g_S^2+{\cal O}(v_L^2/v_R^2)]
\label{nbmass}
\eeq
and eigenvectors
\beq
\left(\begin{array}{c}
\gamma \\ Z \\ Z' \end{array}\right)=
\left(\begin{array}{ccc}
s_W & s_W & \sqrt{c_{2W}} \\
c_W & -t_Ws_W & -t_W\sqrt{c_{2W}} \\
0 & -\sqrt{c_{2W}}/c_W & t_W \end{array}\right)
\left(\begin{array}{c}
W_L^0 \\ W_R^0 \\ B_0 \end{array}\right).
\eeq
For $v_L\ll v_R$,
the usual electroweak relation $M_W^2=M_Z^2 c_W^2$ is still maintained
where we define
\beq
g_L^2=\frac{e^2}{s_W^2} \Rightarrow s_W^2=\frac{3 g_R^2 g_S^2}
                         {3g_R^2 g_S^2+2g_L^2 g_R^2+3g_L^2 g_S^2}.
\eeq

For the scalars we make the simplifying assumption that the different
generations don't mix. In this case, within each generation,
we will get a $2 \times 2$ mass matrix for each of the differently
charged scalars. As an example of the notation we will use, consider
the scalars $L_u^3,~R_u^3$. Diagonalizing their mass matrix will give
us two physical states with masses $M_{u3},~M_{u3}'$
and a mixing angle $s_{u3}$. The masses will generally be of order
$v_R$ and the mixing angles of order $v_L/v_R$. Care should be taken
to spot the scalars that get absorbed to form the longitudinal
components of the massive gauge bosons, and to treat them
appropriately. Since we do our calculation in 't Hooft-Feynman gauge,
these scalars are simply assigned the mass of the corresponding gauge
bosons.

Having enumerated the physical neutrinos, gauge bosons, and scalars,
we can finally list the formulas used to calculate the fermion
masses (the indices ($a,b$) in the following formulas are generation
indices).  For the up-type quarks we get from Fig.~5(a):
\beq
m^{ab}=
2\frac{g_S^2}{(4\pi)^2}\sum_{i=1}^9
U_{a,i}U_{b+3,i} m_i[\frac{m_i^2}{m_i^2-M_X^2}
                                  \ln(\frac{m_i^2}{M_X^2})].
\label{mgu}
\eeq
{}From Fig.~5(b) we get:
\beqa
m_j^{ab} & ={\displaystyle{
\sum_{m,n=1}^3\frac{\kappa_{Lj}^{am}\kappa_{Rj}^{bn}}{(4\pi)^2}
s_{uj} \sum_{i=1}^9 U^*_{m+6,i}U^*_{n+6,i}
m_i}} \nonumber \\
 & {\displaystyle{[\frac{M_{uj}^2}{m_i^2-M_{uj}^2}
\ln(\frac{m_i^2}{M_{uj}^2})
-\frac{M_{uj}'^2}{m_i^2-M_{uj}'^2}
\ln(\frac{m_i^2}{M_{uj}'^2})] }},
\label{msu}
\eeqa
and we sum over $j=1,2,3$ for the three generations of scalars.

For the down-type quarks, Fig.~5(c) gives us
\beqa
m_j^{ab}&={\displaystyle{
\sum_{m,n=1}^3\frac{\kappa_{Lj}^{am}\kappa_{Rj}^{bn}}{(4\pi)^2}
s_{dj} \sum_{i=1}^9 U^*_{m+6,i}U^*_{n+6,i}
m_i}}\nonumber \\
& {\displaystyle{[\frac{M_{dj}^2}{m_i^2-M_{dj}^2}
\ln(\frac{m_i^2}{M_{dj}^2})
-\frac{M_{dj}'^2}{m_i^2-M_{dj}'^2}
\ln(\frac{m_i^2}{M_{dj}'^2})] }}.
\label{msd}
\eeqa
%

For the charged leptons, from Fig.~5(d), we get
\beqa
m_j^{ab}&={\displaystyle{
\sum_{m,n=1}^3\frac{\kappa_{Lj}^{am}\kappa_{Rj}^{bn}}{(4\pi)^2}
s_{ej} \sum_{i=1}^9 U^*_{m+6,i}U^*_{n+6,i}
m_i}}\nonumber \\
& {\displaystyle{[\frac{M_{ej}^2}{m_i^2-M_{ej}^2}
\ln(\frac{m_i^2}{M_{ej}^2})
-\frac{M_{ej}'^2}{m_i^2-M_{ej}'^2}
\ln(\frac{m_i^2}{M_{ej}'^2})] }}.
\label{mse}
\eeqa
Here, however, there is no contribution for $j=1$ since the
$L_e^1$ and $R_e^1$ form the longitudinal components of the $W_L$ and
$W_R$, respectively, and don't mix at tree level.

For the neutrinos, Fig.~5(e) gives us the following Dirac mass matrix
\beq
m^{ab}=
\frac{g_1g_2}{(4\pi)^2}\sum_{i=1}^9
U_{a,i}U_{b+3,i} m_i[\frac{m_i^2}{m_i^2-M_Z'^2}
                                  \ln(\frac{m_i^2}{M_Z'^2})].
\label{mgnd}
\eeq
where
\beq
g_1=\frac{3g_S}{\sqrt{6}};~~g_2=\frac{3t_Wg_S}{\sqrt{6}}+
\frac{g_L\sqrt{c_{2W}}}{c_W}.
\eeq
Fig.~5(f) leads to the following Majorana mass matrix for the
left-handed neutrinos,
\beqa
m^{ab}&={\displaystyle{
2\frac{g_1^2}{(4\pi)^2}\sum_{i=1}^9
U_{a,i}U_{b,i} m_i[\frac{m_i^2}{m_i^2-M_Z'^2}
                            \ln(\frac{m_i^2}{M_Z'^2})]}}\nonumber \\
& + {\displaystyle{
2\frac{g_0^2}{(4\pi)^2}\sum_{i=1}^9
U_{a,i}U_{b,i} m_i[\frac{m_i^2}{m_i^2-M_Z^2}
                                  \ln(\frac{m_i^2}{M_Z^2})]}}
\label{mgnm1}
\eeqa
where
\beq
g_0=\frac{g_L}{c_W}.
\eeq
There will be a similar Majorana mass matrix for the right-handed
neutrinos, $N_a$, with $g_1\rightarrow g_2$, and $a,b\rightarrow
a+3,b+3$ in the first term of Eq.~(\ref{mgnm1}), and no contribution
from the second term. These $3\times 3$ mass matrices for the
neutrinos fill in the appropriate zero blocks
in the $9 \times 9$ mass matrix of
Eq.~(\ref{n9mass}), which is then rediagonalized, resulting in small but
finite see-saw suppressed masses for the left-handed neutrinos.


\end{document}